\documentclass[mathleft,fleqn,%
]{an}
\usepackage{amsmath}
\usepackage{graphicx}
\usepackage[varg]{txfonts}

\usepackage{natbib}
\bibpunct{(}{)}{;}{a}{}{,}
\begin{document}

\Pagespan{1}{}
\Yearpublication{2017}%
\Yearsubmission{2017}%
\Month{}%
\Volume{}%
\Issue{}%
\DOI{}%

\title{Identification spectra of several Northern Hemisphere CV candidates}

\author{H. Worpel\inst{1}\fnmsep\thanks{Corresponding author:
        {hworpel@aip.de}}
\and A. D. Schwope\inst{1}
}
\titlerunning{CV Candidates}
\authorrunning{H. Worpel, A. D. Schwope}
\institute{
Galaxies Group, Leibniz-Institut f\"ur Astrophysik Potsdam (AIP), An der Sternwarte 16, Potsdam, Germany}

\received{}
\accepted{}
\publonline{}


\abstract{%
We obtained Calar Alto identification spectra for six cataclysmic variable candidates, and studied archival observations at a range of wavelengths. Two sources were too faint to allow
for easy identification by their spectra, and the other four are likely to be dwarf novae. No periodicity was detected for any of the sources on the basis of CRTS data.
}

\maketitle

\section{Introduction}

Cataclysmic variables (CVs) are binary stellar systems consisting of a white dwarf primary accreting from
a Roche-lobe filling companion, typically a main sequence star, brown dwarf, or subgiant \citep{Warner1995}.
These systems may accrete through a disc or, in highly magntised systems, by the coupling of infalling material
onto the magnetic field lines of the primary.

CVs are interesting for many reasons. They show variability at all time scales and in all wavelengths, displaying both periodic and aperiodic
brightness changes. They yield important insights regarding the accretion process, with and without discs, in a wide range of physical conditions. Through novae, occasional
unstable explosive episodes of nuclear burning on the surface of the primary, they expel the ashes into space, making them important contributors to the interstellar medium.

Accordingly, CVs are an area of intense research. The identification, classification, and extended monitoring of these
systems remains an important topic in astronomy. Various research groups and projects contribute to this effort,
in different ways. With the Sloan Digital Sky Survey (SDSS; \citealt{StoughtonEtAl2002}) numerous CV candidates were identified
on the basis of their emission lines and led to the discovery of hundreds of CVs \citep{SzkodyEtAl2011}. 
Careful study of X-ray bright point sources have also yielded many promising new candidates \citep{DenisenkoSokolovsky2011}.

Transient searches such as the Catalina Real-Time Transient Survey (CRTS; \citealt{DrakeEtAl2009}), the All-Sky Automated Survey for Supernovae (ASASSN; \citealt{DavisEtAl2015}), and the \emph{Swift}
X-ray telescope discover new candidates by their brightness variations. Such surveys discover a very large number of candidates, but these require followup observations to determine their nature. In this paper
we report on follow-up spectroscopy of six CV candidates identified in the CRTS, performed at the 2.2\,m telescope at the Calar Alto observatory.

\subsection{The sources}
Three of the sources are only known as CRTS transients \citep{DrakeEtAl2014, DrakeEtAl2014b} and have not been otherwise observed or studied. These are
CRTS J072144.5+663838, CRTS J131514.4+424747, and CRTS J162209.6+360419. Another, CRTS J050253.1+171041, is also a CRTS transient and has an unpublished spectrum \citep{DrakeEtAl2014b}.
The other two sources have some previous published observations:

\emph{ CRTS J105122.8+672528}

This object is believed to be an SU UMa dwarf nova with a 0.0596 day period \citep{PavlenkoEtAl2012}, though a later observation
showed no optical variability over a 3.5 hour run \citep{SokolovskyEtAl2012}. The system shows the superhumps
characteristic of SU UMa dwarf novae \citep{KatoEtAl2013}. Possible X-ray and radio counterparts have been identified \citep{TiurinaEtAl2012}.
This source may be associated with the ROSAT source 2RXS J105120.2+672551 \citep{BollerEtAl2016}, some 25" from the optical position. That
X-ray source has a HR1 hardness ratio of 0.68 \citep{VogesEtAl1999}.

\emph{DDE35 = 1RXS J105503.5+681208}

This X-ray source was discovered with ROSAT, with a hardness ratio of 1.0 \citep{VogesEtAl1999}. It is a suspected CV based on long term
photometry \citep{Denisenko2015}, and initial spectroscopy suggests that it is a dwarf nova \citep{FraserEtAl2016}.
It does not appear in the CRTS catalogue.

\section{Methods}

We obtained spectra for our targets on the night of 2016 January 9 (MJD 57396-57397) with the 2.2\,m telescope at Calar Alto. The instrument was
equipped with the Calar Alto Faint Object Spectrograph (CAFOS), a low resolution grism spectrograph and imager. We used the G200 grism, providing a wavelength coverage of 3750\AA\ to
10500\AA\ (though the usable range in our resulting spectra is limited to about 4,000\AA\ to 9,000\AA) at a full width half maximum (FWHM) resolution of about 13 \AA , as measured from arc lamp spectra, through a 1.5 arcsecond slit. The observation log is given in Table \ref{tab:CalarAltoObsLog}.

\begin{figure*}
   \includegraphics{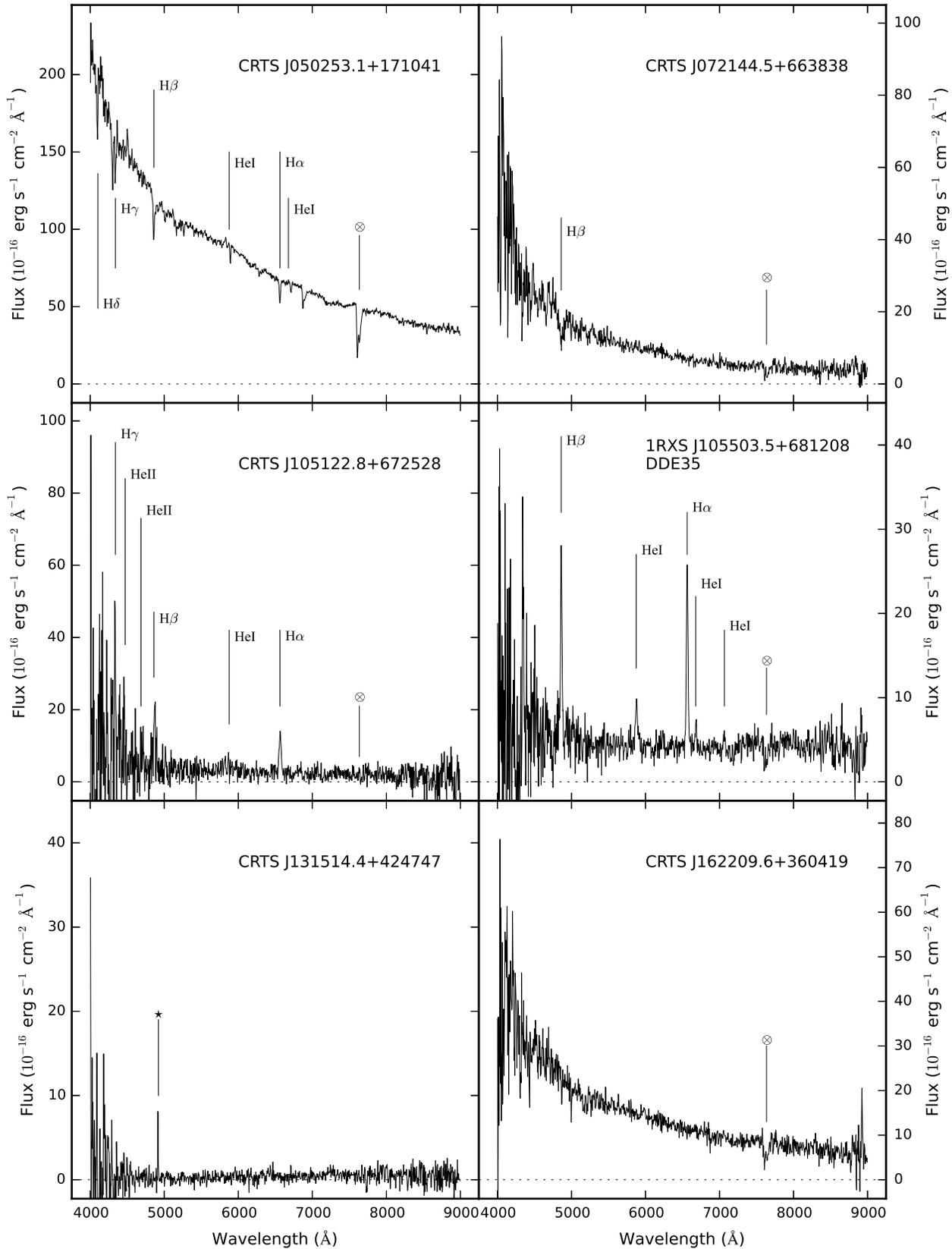}
   \caption{ Spectra for the six sources. If more than one exposure was taken of a source, its average spectrum is shown. Identified spectral lines are labelled, and the uncorrected
             telluric absorption line is marked with a cross symbol, and a cosmic ray with a star. The bottom two sources did not show any obvious lines associated with CVs.}
   \label{fig:spectra}
\end{figure*}

Arc lamp spectra (Hg+He+Rb) for wavelength calibration and standard stars for photometric calibration were taken before and after the
observations of the target stars. The data were reduced and analysed with the ESO-MIDAS software \citep{Warmels1992}. The spectra are shown in Figure \ref{fig:spectra}.

We did not place a second star on the slit. Furthermore, seeing conditions were poor ($\sim 4``$) during the observing run due to strong wind. For 
these reasons we could not easily correct for non-photometric conditions. Our analysis is therefore reliable only for the shape of the spectra, and the presence
or absence of emission lines. We estimate a 30\% uncertainty or 0.3 magnitudes in the overall photometric brightness; good enough to determine whether the objects were in quiescence or outburst.

We calculated approximate V-band magnitudes for our spectra by integrating the flux spectra between 5000\AA\ and 5900\AA\ and comparing these to the standard star G191B2B, correcting for varying airmass because the observations were not at the same time. We estimated the atmospheric extinction coefficient for Calar Alto at 5450\AA\ to be 0.115, by interpolation of the values for winter given in \cite{SanchezEtAl2007}, Table 5. We list the results in Table \ref{tab:CalarAltoObsLog}.

\begin{table}
   \caption{ Calar Alto observation log (MJD 57396-57397)}
   \begin{tabular}{rcrcc}
      Target & Exposures & Length & V mag & Outburst?\\
      \hline
      J0502 & 3 &  600\,s & 15.7 & Yes\\
      J0721 & 3 &  600\,s & 17.9 & Yes \\
      J1051 & 1 &  600\,s & 19.4 & No \\
      DDE35 & 5 &  600\,s & 19.0 & No \\
      J1315 & 1 & 1200\,s & 22.5 & No \\
      J1622 & 1 & 1200\,s & 17.6 & Yes \\
   \end{tabular}
   \label{tab:CalarAltoObsLog}
   
\end{table}

\begin{figure}
 \includegraphics{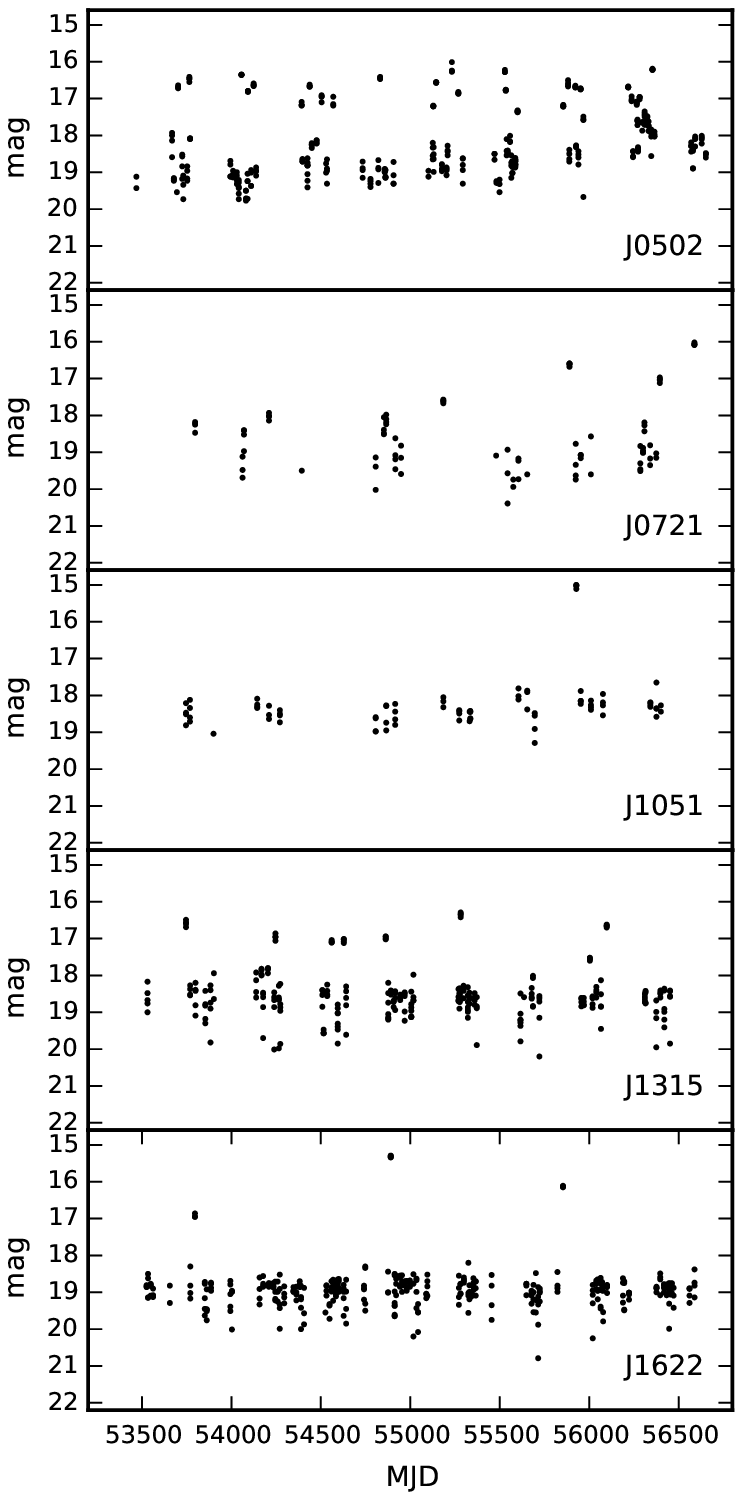}
 \caption{CSS light curves for the CRTS candidates. All show an approximately flat base brightness and at least one outburst of about {1-2} magnitudes.}
 \label{fig:css_licus}
\end{figure}

Five of the six targets (all but DDE35) have long-term photometric data from the Catalina Sky Survey (CSS; \citealt{DrakeEtAl2009}). We downloaded all the relevant data and corrected the timings
to the Solar System barycenter using the algorithms provided by \cite{EastmanEtAl2010}. We sought spin and orbital periods using the Analysis-of-Variance (AoV) method \citep{Schwarzenberg-Czerny1989}. In 
some cases the magnitudes were bimodally distributed with a separation of 1-2 magnitudes, consistent with either a magnetic CV containing bright spots that rotate in and out of view, or a dwarf nova
exhibiting irregularly timed outbursts. We performed the period searches between ten minutes and twelve hours, both including and excluding the brighter data points, but there was no
significant periodicity found for any of the targets. The CSS light curves for these objects are given in Figure \ref{fig:css_licus}.

None of the targets have archival X-ray observations by \emph{XMM-Newton} or \emph{Swift}, and only DDE35 and possibly J1051 were detected by ROSAT.

\begin{figure*}
 \includegraphics{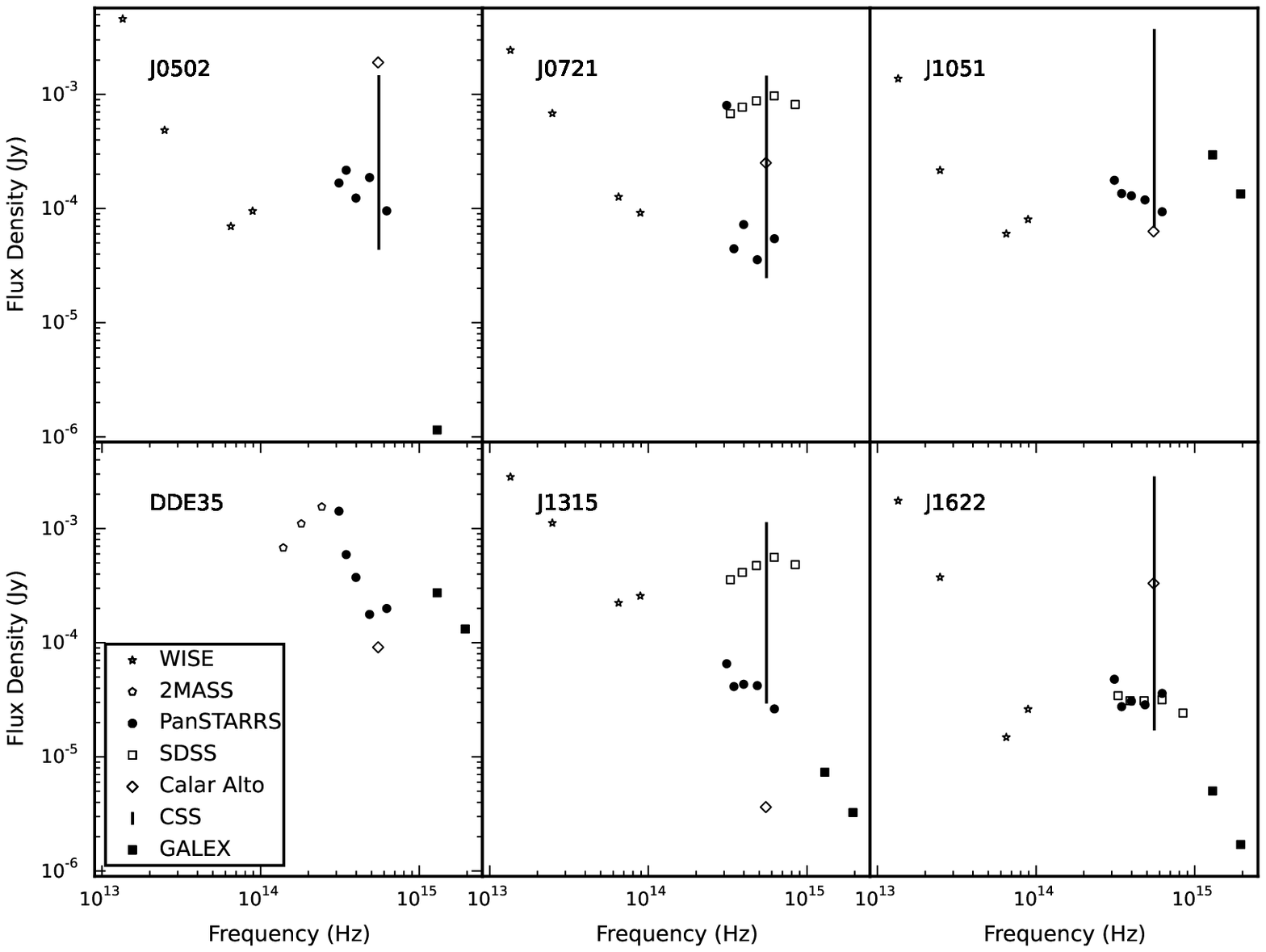}
 \caption{Spectral energy distributions for the six targets. The observations were not simultaneous so intrinsic variability is possible. The vertical bar for the CSS data represents the entire
          range of magnitudes spanned by the CSS observations.}
 \label{fig:seds}
 
\end{figure*}

To obtain a better understanding of the spectral energy distributions (SEDs) of our targets, we looked for archival observations in PanSTARRS, the
Sloan Digital Sky Survey, GALEX, 2MASS, and WISE data. We then constructed SEDs for each source using the effective wavelengths and zero points
given in \citep{TonryEtAl2012, FukugitaEtAl1996, MorrisseyEtAl2007, CohenEtAl2003, JarrettEtAl2011} respectively. We have also included the range spanned
by the CSS data for each source. Since CSS uses unfiltered photometry, we have assumed an effective wavelength of 5,400\AA\ and the same photometric zero point
as the green and red filters in SDSS.  We use that same zero point, and an effective wavelength of 5450\AA\ for the Calar Alto magnitudes. The results are shown in Figure \ref{fig:seds}.

We see clear evidence for outbursts in the sources J0721 and J1315, where in both cases SDSS observed the star in a bright state and PanSTARRS observed it during quiescence.
For DDE35 and J1051, two peaks in the spectral energy distributions are visible. At lower frequencies we may be seeing the contribution of the red secondary star and at higher
frequency the material accreting onto the primary. However we see no evidence for TiO absorption features in their spectra, or an increase in flux at the long wavelength portion of the spectrum,
both of which we would expect to see if an orange- or red-dwarf companion was contributing much of the flux during this observation.

\section{Results and Conclusions}

J0502 shows an absorption line spectrum, with clear H$\alpha$ and He\,I absorption lines, typical of a dwarf nova during outburst (e.g., \citealt{Warner1995}).
Its CSS photometry is also consistent with a dwarf nova, showing occasional brightness increases of 1-2 magnitudes. The catalog of \cite{BreedtEtAl2014} lists
this source as having four outbursts, but the light curve presented in Figure \ref{fig:css_licus} shows it in a bright state significantly more often.
We therefore consider that this object is likely a dwarf nova.

J0721 shows a relatively featureless, blue spectrum. The only obvious spectral line is an absorption line belonging to H$\beta$. Its SED and outburst behaviour are similar to those of J0502,
suggesting that this source is also a dwarf nova.

CRTS J105122.8+672528 may be a dwarf nova, observed outside of an outburst, on the basis of its emission lines.
There were not enough data points in the CSS light curves to be able to confirm the claimed 80\,min period. It has one recorded outburst in \cite{BreedtEtAl2014}, which can also be clearly seen in our light curve, favouring a dwarf nova identification. 
An alternative possibility is that it is a faint magnetic cataclysmic variable, since it apparently shows HeII in emission and this is more typical of magnetic CVs than dwarf novae, and magnetic CVs with short outbursts after long quiescent periods have occasionally been observed previously (e.g., SDSS~J133309.20+143706.9; \citealt{WorpelEtAl2016}). If this object is the same as 2RXS J105120.2+672551, then the relatively hard X-ray
detection would strengthen the case that it is a magnetic CV.

DDE35 has a flat continuum with obvious H and HeI emission lines. It is also likely to be a dwarf nova, observed between outbursts, confirming the suggestion of \cite{FraserEtAl2016}. The lack of HeII emission lines suggests that the star is not strongly magnetic. This spectrum may be slightly contaminated with light from a nearby bright star, that we have been unable to disentangle. We took five spectra of this object but, given the large uncertainty in photometry, there was no evidence of the source varying in brightness over the observing run.

The SED of DDE35 shows two distinct peaks. These features seem consistent with the contribution of the secondary star at the red end, and the accretion disc at higher energies, but the lack of TiO absorption features makes the interpretation of the red bump as the secondary uncertain.

J1315 was very faint during our observing run; several magnitudes fainter than previous observations in CRTS. No obvious spectral features are present, and the continuum appears to be flat. We are therefore unable to confidently classify this object. The presence of outbursts in the CRTS light curves suggests a possible dwarf nova classification, and the extreme faintness of the object during our observation is indicative of a period of very low accretion rate. We encourage further observations.

J1622 is in outburst. It closely resembles J0502, both in the shape of its SED and its very blue color. It is therefore likely that it is also a dwarf nova, despite lacking any obvious spectral absorption lines.

We have identified four of the objects as probable dwarf novae but the other two stars, J1315 and J1622, were too faint to classify their spectra. However, their CSS light curves show irregularly spaced outbursts of 1-2 magnitudes brighter than in quiescence. This behaviour is typical of dwarf novae, so it is likely that these two objects are dwarf novae.
The lack of periodicity in any of the CSS light curves suggests that the brightness variations are not due to an accretion spot moving in and out of view. It is therefore unlikely that these objects are
magnetic cataclysmic variables, which is further evidence of them being dwarf novae. An alternative explanation for J1051, however, is that it is a magnetic CV. We encourage further observations of J1051 and J1315 in particular.

\bibliographystyle{an}
\bibliography{bibli}

\acknowledgements
This work was supported by Deutsches Zentrum für Luft- und Raumfahrt under contracts 50 OR 1405 and 50 OR 1711. This work made use
of data collected at the Calar Alto observatory. The CSS survey is funded by the National Aeronautics and Space
Administration under Grant No. NNG05GF22G issued through the Science
Mission Directorate Near-Earth Objects Observations Program.  The CRTS
survey is supported by the U.S. National Science Foundation under
grants AST-0909182 and AST-1313422. We are grateful to the anonymous referee, whose suggestions led to substantial improvements in the clarity of this paper.

\end{document}